# Airborne Ouzo: Evaporation-induced emulsification and phase separation dynamics of ternary droplets in acoustic levitation

Misaki Mitsuno[1](光野 海祥) and Koji Hasegawa[2,a] (長谷川 浩司)

[1] *Graduate School of Engineering, Kogakuin University, Tokyo, 163-8677, Japan.*

[2] *Department of Mechanical Engineering, Kogakuin University, Tokyo 163-8677, Japan*

## ABSTRACT

We experimentally investigated the evaporation dynamics of acoustically levitated Ouzo droplets (a mixture of ethanol, water, and anise oil). Acoustic levitation has gained significant attention in various fields due to its potential to create a lab-in-a-drop. Although evaporation is a key process in nature and industry, many studies have focused on single and binary components, and ternary droplets in acoustic levitation have rarely been experimentally investigated. In this study, the evaporation-induced spontaneous emulsification (the Ouzo effect) and phase separation process at 40-90 vol% ethanol was visualized. We estimated the concentration change by evaporation of each component in the levitated ternary droplets based on the evaporation model to identify the experimental results. Our experimental results revealed four distinct stages of evaporation in levitated Ouzo droplets: (1) preferential evaporation of the volatile component (ethanol), (2) spontaneous emulsification (myriad micro-oil droplets generation), (3) phase separation forming a core-shell droplet, and (4) water evaporation completion resulting in remaining an oil droplet. Finally, we analyzed the emulsification process by a spacetime diagram. These findings suggest that acoustic levitation is suitable for physicochemical manipulation in mid-air.

\_\_\_\_\_\_\_\_\_\_\_\_\_\_\_\_\_\_\_\_\_\_\_\_

[a]Authors to whom correspondence should be addressed: Electronic mail: kojihasegawa@cc.kogakuin.ac.jp

## I. INTRODUCTION

Contactless fluid manipulation has been extensively explored in analytical chemistry[1–3], biology[4–6] and materials science[7–9]. The acoustic levitation method is a powerful and promising technology because it can levitate any sample (regardless of the thermophysical properties of the solid and liquid) in mid-air using a resonant acoustic field, allowing for lab-in-a-drop processes (droplet injection, levitation, transportation, coalescence, mixing/reaction, evaporation, and ejection).[10–15] Acoustic levitation can prevent undesirable wall effects, such as nucleation and contamination due to container walls. However, the strong acoustic fields used to levitate the sample produce nonlinear effects on the levitated droplet, such as acoustic streaming[16–20] and dynamic interfacial behavior[21-24]. These phenomena greatly affect the evaporation dynamics of the levitated samples. In a theoretical study, Yarin et al. found that the flow occurring around levitated droplets affects the mass transport[25]. Hasegawa et al. studied the correlation between the flows inside and outside the levitated binary droplet under the evaporation process.[26] Bänsch et al. studied the evaporation behavior of levitating droplets by using numerical simulations, considering the deformation of the droplets because of the heat and mass transport at the droplet interface, acoustic flow, and acoustic radiation pressure.[18] Zaitone proposed a theoretical model for the evaporation of spheroidal droplets, considering the effect of the acoustic flow.[27] For the binary droplet, Yarin et al. developed an evaporation model for multicomponent droplets levitated in an acoustic field.[28] Niimura et al. observed the evaporation process of water-ethanol mixture droplets, and the existing theory was extended for the evaporation of binary droplets based on the experimental results.[29] Regarding the nucleation dynamics associated with the evaporation of multicomponent droplets, Chen et al. observed the liquid-phase separation processes that occur during the evaporation, using binary droplets.[30] For Ouzo effect, Tan et al. investigated the sessile ternary droplet can trigger a phase transition and the nucleation of microdroplets of one of the components of the mixture novel visualization by the confocal microscope.[31]

While many studies have explored the evaporation process of pure and binary components using acoustic levitation, few have investigated the evaporation and nucleation dynamics in multicomponent droplets containing three or more components. Additionally, there has been limited research on levitated Ouzo droplets. In this study, we aimed to understand the evaporation dynamics associated with the spontaneous emulsification and phase separation of Ouzo droplets premixed with ethanol, water, and anise oil. Therefore, the spontaneous emulsification and phase separation processes of Ouzo droplets during evaporation were visualized and analyzed using a droplet evaporation model. The droplet levitation dynamics with evaporation-induced emulsification were presented to facilitate a more universal understanding for the development of future lab-in-a-drop applications.

## II. EXPERIMENTAL DESIGN

### A. Experimental setup

Figure 1 (a) shows a schematic diagram of the experimental setup used in this study[14]. A sinusoidal signal generated by a function generator was input to the ultrasonic transducer. The ultrasonic wave is emitted from the bottom horn connected to the ultrasonic transducer and is reflected by the top reflector, which is placed at a multiple of half the wavelength distance to form an acoustic standing wave between the horn and the reflector, as shown in Fig. 1 (b). To levitate the droplet, a liquid droplet was manually injected using a syringe and needle near the acoustic pressure node of the acoustic standing wave. A high-speed camera (Photron, FASTCAM Mini AX200) with backlight illumination was used to visualize the levitated droplets. The interface temperature of the levitated droplet was simultaneously captured using an IR camera (FLIR Systems, A6750sc, MWIR). The images obtained by the high-speed camera were processed using a computer and in-house code (MATLAB R2023a). The reflector used in this experiment had a radius of curvature of 36 mm (R36). The frequency is 19.3 kHz and the wavelength $\lambda$ is 18 mm. The width of the horn and reflector is 36 mm (= $2\lambda$). The distance between the horn and the reflector was 48 mm (≈ 5λ/2), and a droplet was levitated near the third pressure node (≈ 3λ/2 from the bottom reflector). Experiments were conducted at room temperature (23±2°C) and humidity (55±5%). The sample consisted of a three-component liquid premixed with varying concentrations of ethanol, water, and anise oils to observe the Ouzo effect. The initial equivalent diameters of the droplets d ranged from 1.46 mm to 2.22 mm, and the initial aspect ratio $AR$ was between 1.0 and 2.8. $AR$ $(=b/a)$, which was defined as the equatorial-to-polar ratio of the major diameter $b$ and minor diameter $a$. The droplet diameter and aspect ratio were quantified using image analysis.

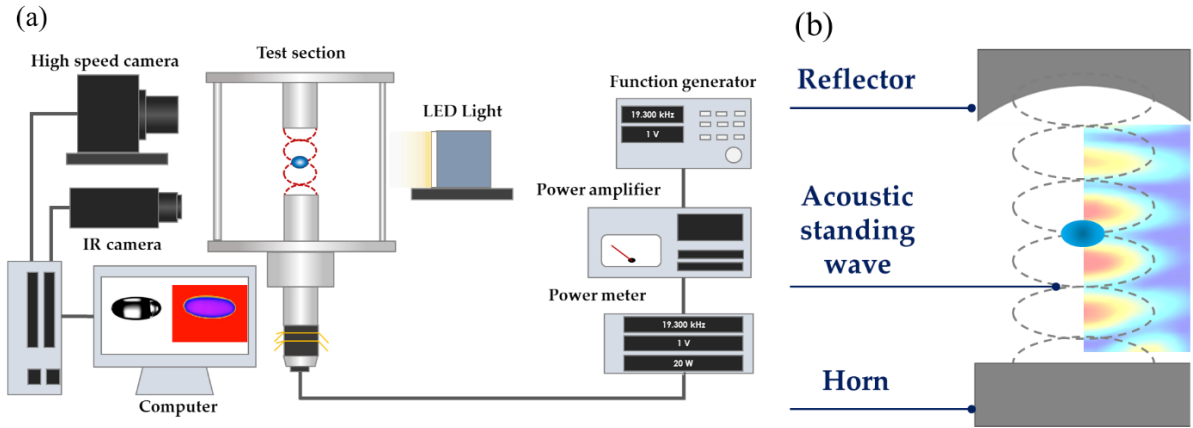

FIG. 1. Schematic of acoustic levitator: (a) experimental setup and (b) levitation principle.

Table I lists the thermophysical properties of the pure liquid. The importance of solubility in the Ouzo effect on the levitated droplets must be highlighted. The premixed droplet becomes soluble because of the solubility of ethanol in both water and anise oils. The emulsification of anise oil, which is immiscible in water, begins in the levitated droplet as ethanol preferentially evaporates and its concentration decreases.

## B. Image analysis: Spacetime diagram

To characterize the emulsification dynamics in the Ouzo droplet, we used a spacetime diagram by an image processing. Figure 2 shows the image processing procedure to obtain the spacetime diagram. Using the images obtained from the experiment, the 1-pixel width slice at the center of gravity of the droplet was extracted from each frame. The extracted slice images were integrated into a spacetime diagram.

TABLE I. Thermophysical properties of test sample at 20°C.[32]

| Fluid | Density $\rho$ [kg/m$^3$] | Diffusion coefficient $D \times 10^{-5}$ [m$^2$/s] | Vapor pressure $p_s$ [kPa] | Solubility |
|---|---|---|---|---|
| Water | 997 | 2.49 | 2.34 | Miscible in in ethanol Immiscible in anise oil |
| Ethanol | 785 | 1.20 | 5.58 | Miscible in ethanol and anise oil |
| Anise oil | 979 | N/A | N/A | Immiscible in water Miscible in in ethanol |

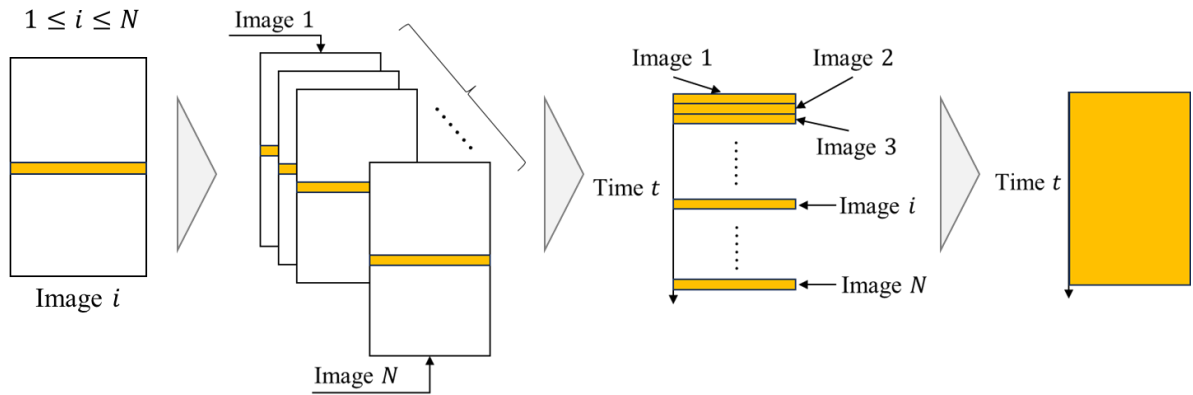

FIG. 2. Procedure to obtain the spacetime diagram.

## III. RESULTS AND DISCUSSION

### A. Evaporation process of Ouzo droplets

Figure 3 shows the evaporation process of the Ouzo droplets with the concentration of 60, 39, 1 vol% of ethanol, water, and anise oil, respectively. Figure 3 (a) demonstrate that the droplet diameter decreased over time and that the Ouzo effect occurred immediately after droplet levitation (~60 s). During the Ouzo effect, the droplets became cloudy (black) because the backlight could not penetrate by the generated myriad micro-oil droplets. Once the Ouzo effect ceased, the micro-oil droplets coalesced due to the internal flow[30] and resulted in core oil droplets in the levitated droplet. For core-shell droplets, although it is natural that the water droplet became the core because of the higher surface tension than that of anise oil, the oil droplet can be the core for an acoustically levitated Ouzo droplet. Figure 3 (b) illustrates the time evolution of the droplet diameter and aspect ratio in Fig. 3 (a). The blue area highlighted in Fig. 3(b) represents the time when each droplet experienced the Ouzo effect, which lasted until approximately 50 s. The experimental results showed that levitated droplet evaporated and nonlinearly decreased the droplet diameter. During the evaporation process, the droplet gradually changed shape and became more spherical ($AR \approx 1$). This nonlinear behavior was due to the preferential evaporation of the volatile ethanol component (~200 s) followed by the evaporation of water (200 s~).[26] After the preferential evaporation of ethanol, the higher surface tension of water maintains the spherical shape of the droplet interface.

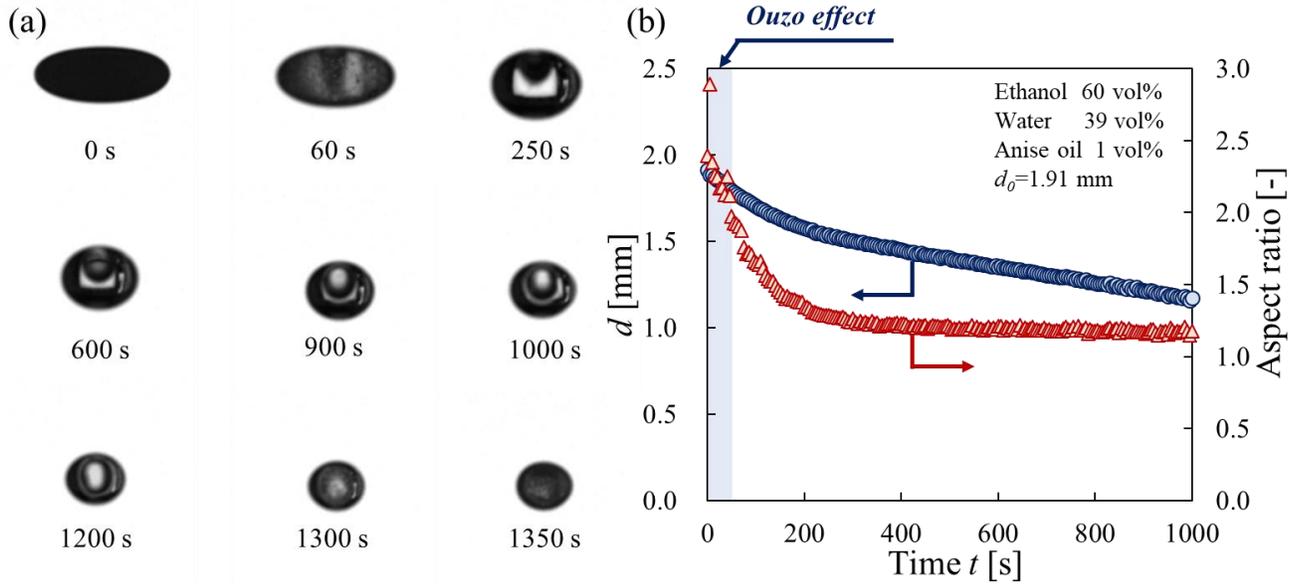

FIG. 3. Evaporation process of Ouzo droplets (ethanol: 60 vol%, water: 39 vol%, anise oil: 1 vol%): (a) Snapshot of Ouzo droplet during evaporation and (b) time evolution of $d$ and $AR$. The highlighted blue area in (b) indicates the occurrence of the Ouzo effect.

Figure 4 (a) shows the evaporation process for ethanol concentrations of 50, 70, and 90 vol% with the different diameter. The concentration of anise oil was constant at 1 vol% in Fig. 4. On the other hand, Fig. 4 (b) displays a normalized surface area $(d/d_0)^2$ as a function of Fourier number. where $d$ and $d_0$ are the instantaneous and initial droplet diameters, respectively. The Fourier number, which characterizes the diffusion phenomenon, can be defined as

$$Fo = \frac{Dt}{d_0^2}, \qquad (1)$$

where $t$ is the time and $D$ is the diffusion coefficient. The diffusion coefficient of ethanol was used in the study. These results illustrate a gradual change in droplet size during evaporation depending on the ethanol concentration. Notably, Fig. 4 (b) shows that the experimental data for evaporating droplets with the same concentration were well characterized by dimensionless values. Particularly, for 70 vol% ethanol, the experimental data successfully collapsed. Thus, we demonstrated that the Fourier number is effective for characterizing the evaporation of levitated Ouzo droplets.

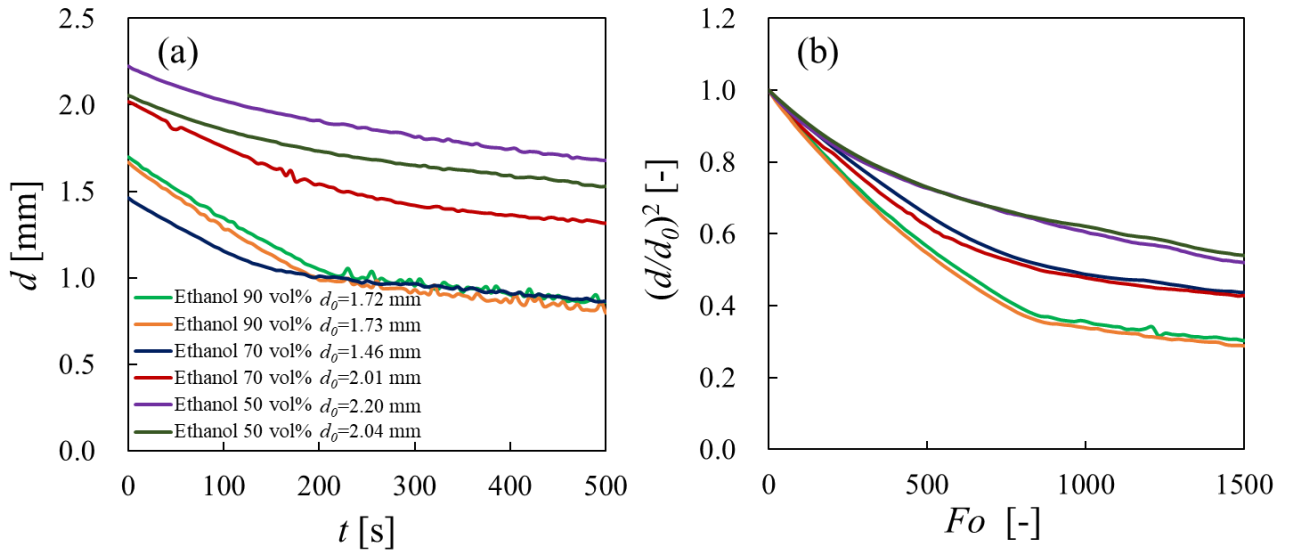

FIG. 4. Evaporation process of Ouzo droplets for different droplet diameter and initial concentration: (a) dimensional form and (b) dimensionless form.

Based on our experimental results, the evaporation of Ouzo droplets in acoustic levitation with four distinct phases is illustrated in Fig. 5: (1) preferential evaporation of ethanol, (2) onset of the emulsification with myriad tiny oil droplets (Ouzo effect), (3) phase separation with the core-shell structure due to coalescence of oil droplets, and (4) evaporation completion resulting in remaining the oil droplet. It is noteworthy that the core-shell structure was observed as shown in Fig. 7 (a). The oil droplet in the water droplet was floated near the top of the droplet interface due to the density difference listed in Table I. To the best of our knowledge, the formation of core-shell structures has not been reported for the sessile Ouzo droplet.[31] These findings can facilitate our understanding of the surfactant-free oil-in-water emulsion in levitated ternary droplets, paving the way for contactless droplet manipulation for an airborne microfluidics. For a deeper understanding of the evaporation and emulsification processes, instantaneous concentration information for the Ouzo effect need to be predicted in the following subsection.

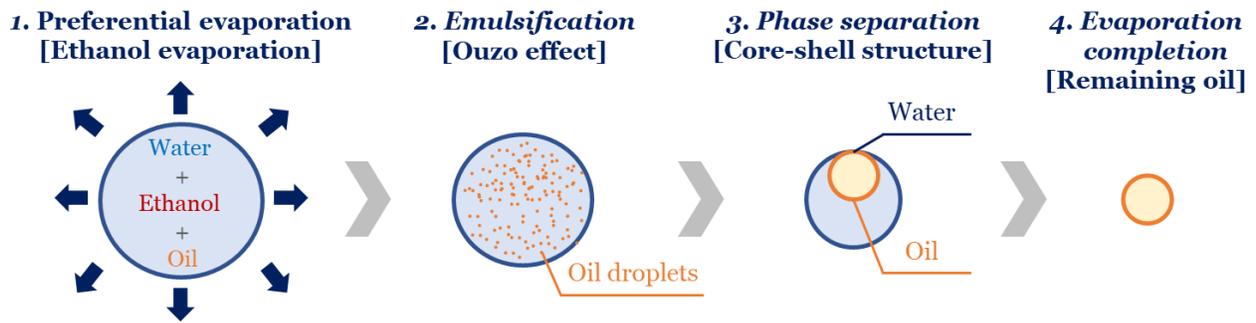

FIG. 5. Evaporation, spontaneous emulsification and phase separation process of Ouzo droplet in acoustic levitation.

## B. Concentration estimation of Ouzo droplets by evaporation model

It can be assumed that the onset and disappearance of the Ouzo effect are due to changes in the concentration inside the droplet. Therefore, it is necessary to explore the instantaneous concentration conditions for the Ouzo effect based on the experimental results and evaporation models. We predicted the concentrations of the ethanol, water, and anise oil components in the levitated droplet. The mass fraction of each component was obtained using Eqs. (2)–(4).

$$Y_e = \frac{m_e}{m_e + m_w + m_A},  \tag{2}$$

$$Y_w = \frac{m_w}{m_e + m_w + m_A},  \tag{3}$$

$$Y_A = \frac{m_A}{m_e + m_w + m_A},  \tag{4}$$

where $Y$ is the mass fraction and $m$ is the droplet mass. The subscript $e$ represents ethanol, $w$ represents water, and $A$ represents anise oil. The masses shown here can be calculated by determining the theoretical droplet diameter using the classical droplet evaporation model (the well-known $d^2$-law[33]). The $d^2$-law can predict droplet evaporation by considering droplet mass transport by diffusion, which is widely used in the estimation of single-droplet evaporation. $d^2$-law is given by Eq. (5).

$$d^2 = d_0^{\,2} - \frac{8DM}{\rho_l R}\left(\frac{p_s}{T_s} - \frac{p_\infty}{T_\infty}\right)t,  \tag{5}$$

where $M$ is the molar mass, $\rho_l$ is the liquid density, $R$ is the gas constant, $P$ is the vapor pressure, $T$ is the temperature, and $t$ is the time. The subscript $s$ represents the droplet surface and $\infty$ represents the ambient gas. This evaporation model was used to obtain the theoretical droplet diameter $d_{th}$, as described by Eqs. (6) and (7). Using the theoretical droplet diameter, the mass of the ethanol component in the evaporating droplet can be estimated using Eq. (8).

$$d_{th} = \sqrt{d_0^{\,2} - \beta t}  \tag{6}$$

$$\beta = \frac{8DM}{\rho_l R}\left(\frac{p_s}{T_s} - \frac{p_\infty}{T_\infty}\right)  \tag{7}$$

$$m_e = \frac{4}{3}\rho_e \pi \left(\frac{d_{th}}{2}\right)^3 \tag{8}$$

The instantaneous droplet mass can be quantified from the experimental data using Eq. (9). The masses of water and nonvolatile anise oil were calculated from each volume obtained from the droplet diameter and density, as expressed by Eqs. (10) and (11), where $V$ is the initial volume of each component. Water vapor from ambient air condenses on the surface of acoustically levitated droplets over time.[29] Meanwhile, we assumed that anise oil is nonvolatile so that the mass of anise oil is constant during the evaporation. Therefore, we estimated the mass of water in the evaporating droplet by calculating the difference between the experimental and theoretical values.

$$m_{exp} = \frac{4}{3}\rho_{exp} \pi \left(\frac{d}{2}\right)^3 \tag{9}$$

$$m_w = (m_{exp} - m_e) + \rho_w V_w \tag{10}$$

$$m_A = \rho_A V_A \tag{11}$$

Figures 6(a)–(c) show the time evolution of the normalized surface area, estimated mass fraction, and surface temperature of Ouzo droplets with the concentration of 70 vol% ethanol, 29 vol% water, and 1 vol% anise oil. As shown in Fig. 6 (a), the droplet evaporated, causing a decrease in the concentration of ethanol due to preferential evaporation, while the concentration of water increased due to vapor condensation in Fig. 6 (b). The temperature difference between the droplet surface and the ambient air, as shown in Fig. 6 (c), promotes the condensation of water vapor around the droplet. To better characterize the Ouzo effect, we discuss the detailed emulsification process in the following subsection.

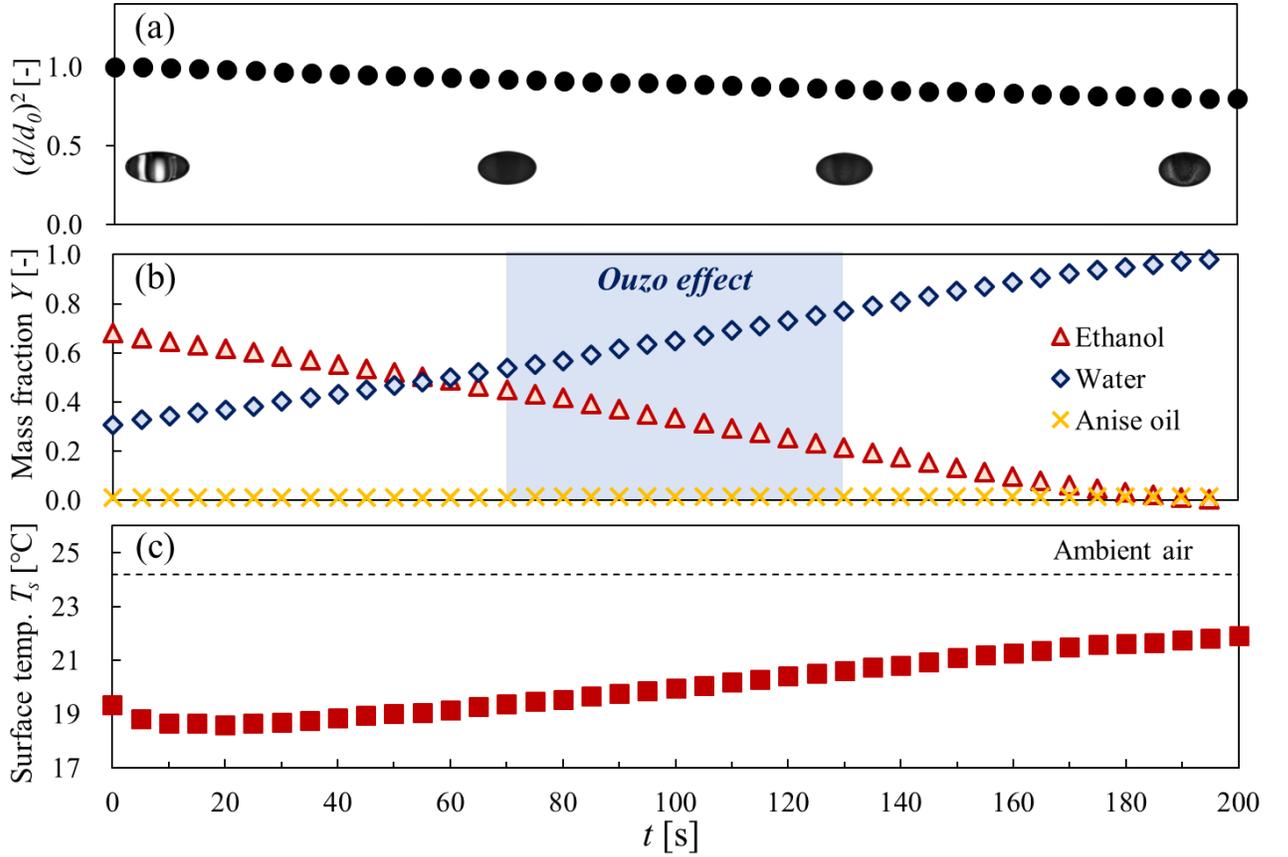

FIG. 6. Time evolution of (a) normalized surface area, (b) mass fraction estimated by Eqs. (2)–(4), and (c) surface temperature obtained by IR camera for Ouzo droplet. (ethanol: 70 vol%, water: 29 vol%, anise oil: 1 vol%). The highlighted blue area in (b) indicates the occurrence of the Ouzo effect.

## C. Emulsification process of Ouzo droplets

Figure 7 shows the characteristic time of Ouzo effect associated with evaporation for the ethanol concentration of 40–90 vol% with 1 vol% anise oil. Figure 7 (a) represents the effect of the ethanol concentration on the start and end time of Ouzo effect. The start time was defined as the moment when emulsification appeared in the droplet, and the end time was when it disappeared, as observed through visualization. Figure 7 (b) shows the effect of the ethanol concentration on the duration of the Ouzo effect. These results indicate that the Ouzo effect occurs during droplet levitation when the ethanol concentration is 40–90 vol%. However, no clear changes were observed in the levitated droplets below 30 vol% ethanol, and the onset of the Ouzo effect could not be confirmed. We also found that higher ethanol concentrations delayed the onset and shortened the duration of the effects of Ouzo. Thus, as the concentration of ethanol increases, the proportion of volatile components in the droplet also increases, resulting in a decrease in the initial water component for the Ouzo effect and a shorter duration.

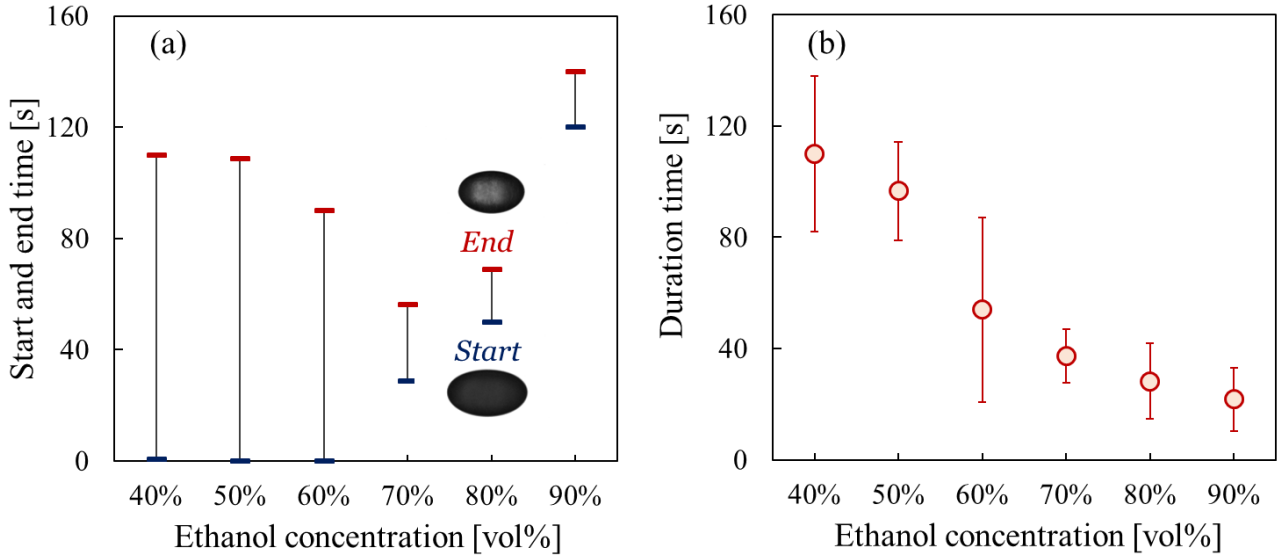

FIG. 7. Characteristic time of Ouzo effect associated with evaporation with different ethanol and water concentration for 1 vol% anise oil: (a) Start and end time of Ouzo effect and (b) duration time of Ouzo effect. Error bar indicates the uncertainty values for each plot.

Figure 8 shows the effect of anise oil concentration on the duration time of the Ouzo effect. The Ouzo effect occurred at ethanol concentrations above 40 vol% and anise oil concentrations up to 20 vol%. We also observed that the duration time decreased as the anise oil concentration increased and the Ouzo effect occurred from the initial state. We believe that the reason for this short duration is the high concentration of anise oil and the rapid separation of oil and water. Although the Ouzo effect was not observed with a 20 vol% anise oil concentration for the sessile Ouzo droplet, it was found for the acoustically levitated droplet. This can be attributed to heat and mass transfer through the liquid-gas interface.

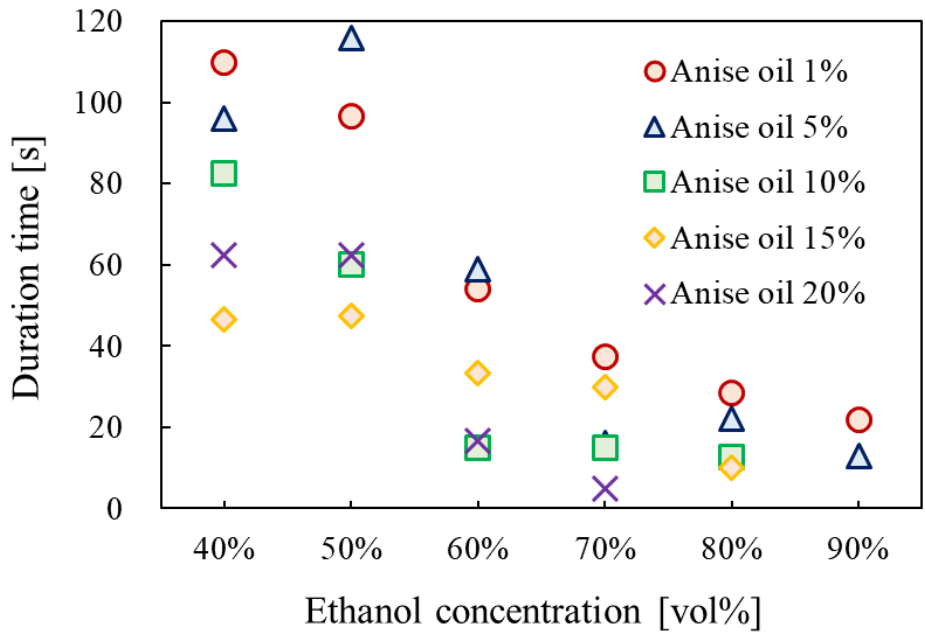

FIG. 8. Effect of anise oil concentration on Ouzo effect. The experimental data for 1% volume of anise oil is identical to that shown in Fig. 7(b).

Figure 9 shows the spacetime diagram of the Ouzo droplets at different ethanol concentrations for 1 vol% anise oil. Figure 9 (a) presents the evaporation and emulsification process of 70 vol% ethanol concentration. The levitated droplet was captured from the top in Fig. 9 (a). From 20 s to 60 s, we can confirm that the Ouzo effect occurs near the droplet interface. This can be due to the oil-water separation as ethanol evaporates. The generated myriad micro-oil droplets start merging after the emulsification.

Figures 9 (b)–(d) show the spacetime diagram for ethanol concentrations of 50, 70, and 90 vol%, respectively. In the color contour, the highlighted yellow and purple indicate the white (bright) and black (dark) in brightness values in the original images. Therefore, a yellow region in each figure represents the occurrence of the Ouzo effect. The spacetime diagram of each droplet revealed that the onset of the Ouzo effect was delayed as the ethanol concentration increased. The duration of the Ouzo effect also decreased with increasing ethanol concentration. These results were consistent with the experimental data presented in Fig. 7. It is hypothesized that although a higher volatile component accelerates the evaporation rate while enhancing the condensation of ambient water vapor, the insufficient initial water amount in the droplet leads to an early termination of the Ouzo effect. This suggests that the concentration of the water component plays a crucial role in the onset and disappearance of the Ouzo effect.

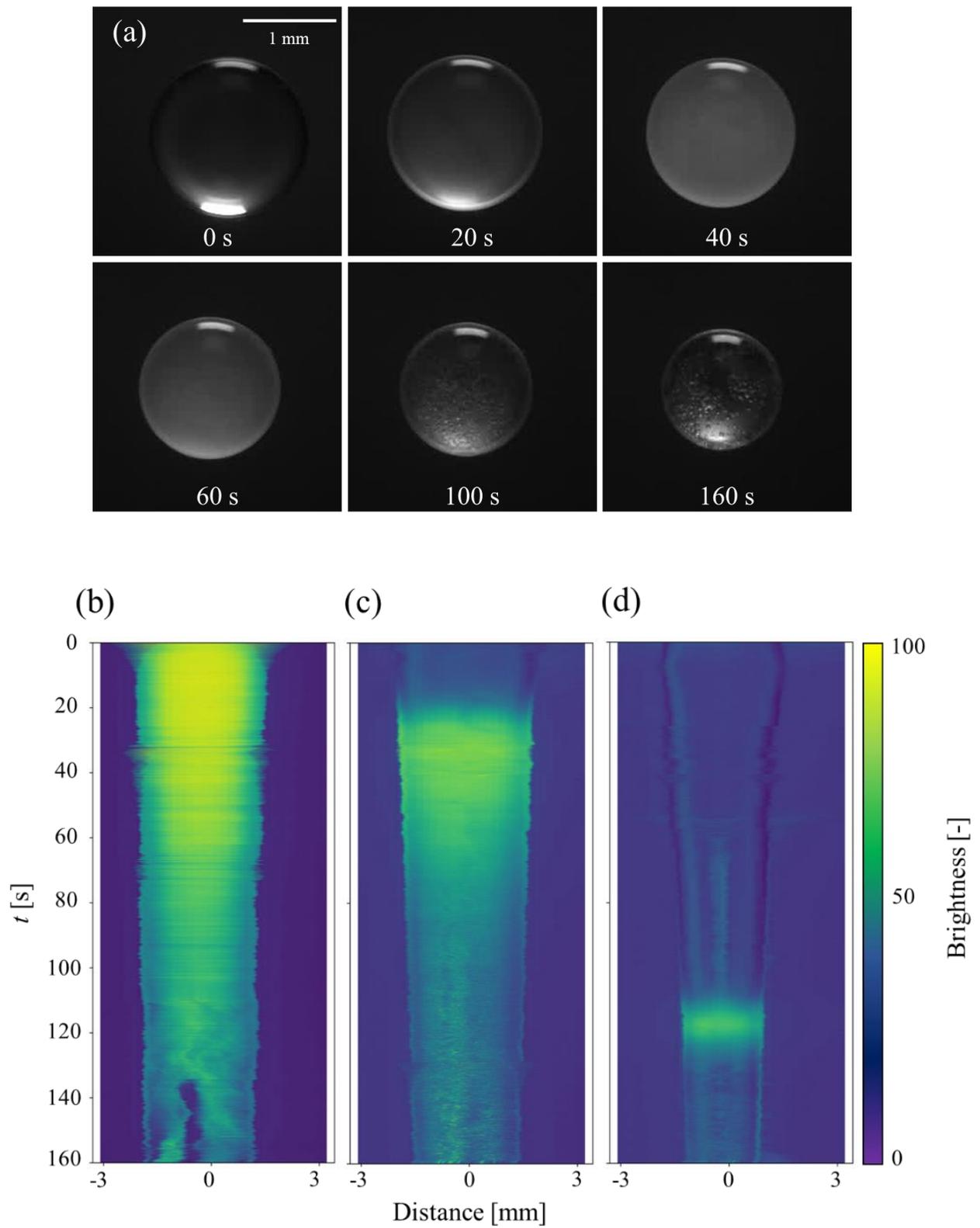

FIG. 9. Spacetime diagram of Ouzo droplets: (a) Original images for an ethanol concentration of 70 vol% and ethanol concentration of (b) 50, (c) 70, and (d) 90 vol% with 1 vol% anise oil.

The time variation in the brightness values obtained from Fig. 9 is depicted in Fig. 10. The results demonstrated that the brightness value increased during the Ouzo effect and decreased once it ended. This was attributed to the formation of myriad oil droplets by the onset of the Ouzo effect, which reflected light and increased the brightness. The inset clearly indicates that the maximum brightness value decreases as the ethanol concentration increases. Thus, we quantified that the emulsification characteristics by brightness values during the Ouzo effect. The number density and size distribution of the generated oil droplets need to be investigated to better understand the emulsification process in an acoustically levitated Ouzo droplet; however, this is beyond the scope of the present study and explored in future studies.

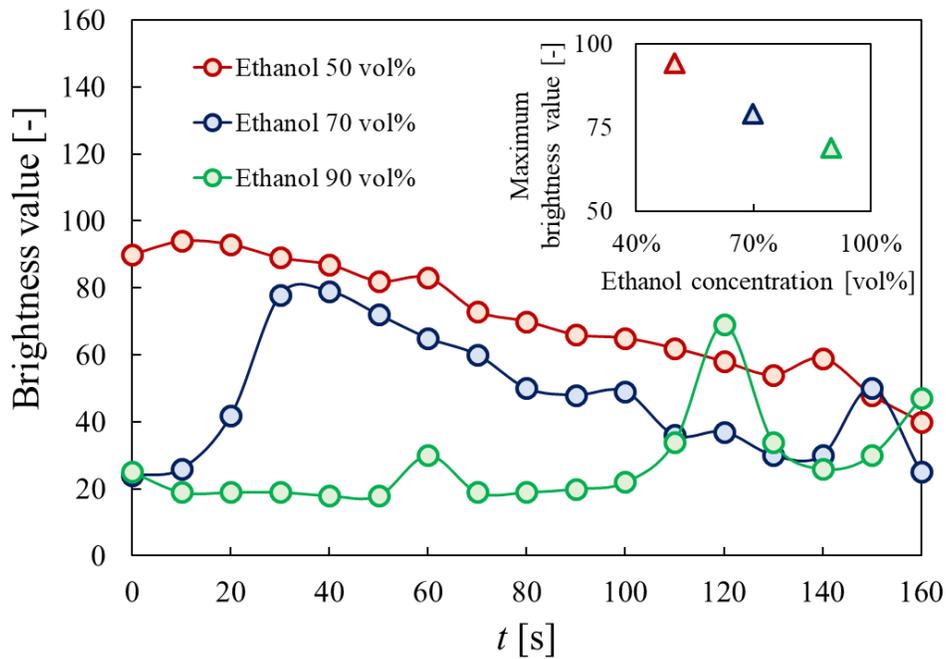

FIG. 10. Time variation of brightness values in Figs. 9 (b)–(d) at the distance of 0 mm.

## IV. CONCLUSIONS

In summary, we demonstrated the evaporation-induced spontaneous emulsification and phase separation dynamics of an acoustically Ouzo droplets using a visualization technique and droplet evaporation model. Our findings indicated that Ouzo effect can be observed even under noncontact conditions by acoustic levitation. The experimental data revealed a four-step process involving the evaporation, spontaneous emulsification, and phase separation of Ouzo droplets in an acoustic field. Our evaporation model well characterized the Ouzo effect induced by the dynamic concentration change in the ternary levitated droplet. Furthermore, the emulsification process was analyzed with varying the anise oil concentrations and clearly visualized by a spacetime diagram. The Ouzo effect occurred between 40 and 90 vol% ethanol. At higher ethanol concentrations, the onset of the Ouzo effect was delayed, and its duration was shorter. To gain a better understanding of the formation process of a core-shell structure in evaporating droplets, future research should focus on quantifying the number density and size distribution of the oil droplets generated in levitated Ouzo droplets. These experimental findings will stimulate further research and be useful for potential lab-in-a-drop applications,[34,35] such as microreactors.[36] Understanding the complex dynamics of multicomponent droplets during acoustic levitation can provide a better knowledge platform for developing rich and diverse applications.

## ACKNOWLEDGEMENTS

This work was supported by JSPS KAKENHI Grant Number 20H02070 and 23K17732. We thank Dr. Xiao Ma for the insightful discussions.

## AUTHOR DECLARATIONS

### Conflict of Interest

The authors have no conflicts to disclose.

## Author Contributions

**Misaki Mitsuno:** Data curation(lead); Formal analysis (lead); Investigation (lead); Methodology (lead); Validation (lead); Visualization (lead); Writing – original draft (lead). **Koji Hasegawa:** Conceptualization (lead); Formal analysis (supporting); Investigation (supporting); Methodology (supporting); Validation (supporting); Visualization (supporting); Funding acquisition (lead); Project administration (lead); Supervision (lead); Writing – review & editing (lead).

## DATA AVAILABILITY

The data that support the findings of this study are available from the corresponding author upon reasonable request.